\documentclass[runningheads]{llncs}

\usepackage{graphicx}
\usepackage{xcolor}
\usepackage[utf8]{inputenc}
\usepackage{multirow}

\let\svthefootnote\thefootnote

\begin{document}

\title{Predicting intelligence based on cortical WM/GM contrast, cortical thickness and volumetry}

\titlerunning{IQ prediction with multimodal information}

\author{Juan Miguel Valverde\inst{1} \and
Vandad Imani\inst{1} \and
John D. Lewis\inst{2} \and
Jussi Tohka\inst{1}}

\authorrunning{J. M. Valverde et al.}

\institute{AI Virtanen Institute for Molecular Sciences, University of Eastern Finland, Kuopio, Finland \\
\email{\{juanmiguel.valverde,vandad.imani,jussi.tohka\}@uef.fi}\\
\and
Montreal Neurological Institute, McGill University, Montreal, Canada\\
\email{jlewis@bic.mni.mcgill.ca}}

\maketitle              

\begin{abstract}
We propose a four-layer fully-connected neural network (FNN) for predicting fluid intelligence scores from T1-weighted MR images for the ABCD-challenge. In addition to the volumes of brain structures, the FNN uses cortical WM/GM contrast and cortical thickness at 78 cortical regions. These last two measurements were derived from the T1-weighted MR images using cortical surfaces produced by the CIVET pipeline. The age and gender of the subjects and the scanner manufacturer are also used as features for the learning algorithm. This yielded 283 features provided to the FNN with two hidden layers of 20 and 15 nodes. The method was applied to the data from the ABCD study. Trained with a training set of 3736 subjects, the proposed method achieved a MSE of 71.596 and a correlation of 0.151 in the validation set of 415 subjects. For the final submission, the model was trained with 3568 subjects and it achieved a MSE of 94.0270 in the test set comprised of 4383 subjects.

\end{abstract}

\let\thefootnote\relax\footnote{J. M. Valverde and V. Imani contributed equally to this work.}
\addtocounter{footnote}{-1}\let\thefootnote\svthefootnote

\section{Introduction}

Fluid intelligence is a core factor of general intelligence. The rate at which skills and knowledge, i.e. crystallized intelligence, are acquired depends upon it. Thus, there is great interest in determining the extent to which fluid intelligence can be determined from brain measures.

In this study, we used a supervised learning model to automatically predict fluid intelligence scores, {\it i.e.} fluid intelligence with demographic confounding factors removed, based on T1-weighted Magnetic Resonance Images (MRIs) at 3T. More specifically, this report describes our submission to the the ABCD Neurocognitive Prediction Challenge (ABCD-NP-Challenge 2019)\footnote{https://sibis.sri.com/abcd-np-challenge/}. Data were obtained from the NIMH Data Archive (NDA) database\footnote{\url{https://nda.nih.gov}}, generated by the Adolescent Brain Cognitive Development (ABCD) study, the largest long-term study of brain development and child health in the United States. The feature set used for the prediction model included sociodemographic (age, gender) and MRI-derived measures. Our MRI-derived features included  regionally averaged cortical thickness and white/grey contrast measures in addition to the volumes of a set of regions of interest provided by the challenge organizers. Relying on this feature set, we create a supervised regression framework utilizing a four-layer fully-connected neural network (FNN) to predict fluid intelligence scores.

\section{Material}

\subsection{Training and Test data}

T1-weighted MR images, volumetric measures, age, gender, scanner and fluid intelligence scores were available to the ABCD challenge participants via National Database for Autism Research (NDAR) website. A primary consideration in measuring general intelligence is the role of fluid intelligence \cite{carroll} which was measured via the NIH Toolbox Neurocognition battery \cite{NIH} and from which demographic factors (e.g., sex and age) are eliminated to remove the effect of confounding variables. For this residualization, the challenge organizers used all subjects without any missing values in the data collection site, sociodemographic variables and brain volume to build a linear regression model. This model was constructed with fluid intelligence as the independent variable and the other attributes as dependent variables. The residuals computed for all subjects provided the fluid intelligence scores to be predicted.

From a total of 8553 individual subjects, fluid intelligence scores were provided for participants in the training set (3736 subjects) and validation set (415 subjects), whereas the other subjects (4402 subjects) formed the test set. As explained in Section 2.2, few of these subjects could not be processed through CIVET pipeline, hence they were not used to predict the fluid intelligence scores. The age and gender characteristics of the 8347 subjects across scanners from three different manufacturers used in this work are presented in Table \ref{tab:table1}. There was no significant difference in age and gender between different scanners.

The FNN was trained with a set of 3568 subjects and validated during the training process on the validation set consisting of 396 subjects. Afterwards, the proposed method was used to generate the predictions of the fluid intelligence scores of the 4383 subjects in the test set.

\begin{table}
\begin{center}
\caption{The subject characteristic of this work. The ages of subjects ranged from 107 to 133 months.}\label{tab1}
 \begin{tabular}{|l@{\hskip 0.05in}|c|c|c|c|c|c|c|c|c|c}
    \hline
    \multirow{2}{*}{Scanners} & \multicolumn{3}{c|}{GE (25\%)} &\multicolumn{3}{c|}{SIEMENS (60\%)} &\multicolumn{3}{c|}{Philips (15\%)} \\\cline{2-10} 
    
        & \hspace{0.01in} Train \hspace{0.01in} & \hspace{0.01in}Validation\hspace{0.01in} & \hspace{0.01in}Test\hspace{0.01in} & \hspace{0.01in}Train\hspace{0.01in} & \hspace{0.01in}Validation\hspace{0.01in} & \hspace{0.01in}Test\hspace{0.01in} & \hspace{0.01in}Train\hspace{0.01in} & \hspace{0.01in}Validation\hspace{0.01in} & \hspace{0.01in}Test\hspace{0.01in}\\
    \hline
    Subjects & 966 & 99 & 1029 & 2101 & 240 & 2698 & 501 & 57 & 656 \\
    \hline
    Females & 48.2\% & 47.4\% & 49.1\% & 46\% & 51.6\% & 47.3\% & 49\% & 45.6\% & 49.2\% \\
    \hline
\end{tabular}
\label{tab:table1}
\end{center}
\end{table}
 
\subsection{Image pre-processing}

Volume measures, provided by the competition organizers, were derived from the T1-weighted images as follows: the Minimal Processing pipeline \cite{minimalprocessingpipeline} transformed the raw data into NIfTI format. Afterwards, the NCANDA pipeline \cite{ncandapipeline} defined a brain mask by non-linearly mapping the SRI24 atlas \cite{SRI24} to the T1-weighted images, and it removed noise and corrected for bias-field inhomogeneities. Several skull-stripping methods were used with bias-field corrected and non-bias-field corrected images, and a majority voting of the resulting masks refined the brain masks from the previous step. The skull-stripped brains were newly corrected for bias-field inhomogeneities and segmented into gray matter, white matter and cerebrospinal fluid via Atropos \cite{atropos}. The SRI24 atlas was non-rigidly registered to the images via ANTS \cite{ants} to further parcellate the gray matter and the resulting segmentations were linearly registered to the SRI24 atlas. Finally, results that failed to pass a visual two-tier quality check were rejected.

In addition to volume measures, we used cortical thickness and cortical white/gray contrast measures that were regionally averaged based on the Automated Anatomical Labeling (AAL) atlas \cite{tzourio2002automated}. For this, the T1-weighted volumes were denoised \cite{manjon2010adaptive} and processed with CIVET (version 2.1 ; 2016), a fully automated structural image analysis pipeline developed at the Montreal Neurological Institute\footnote{\url{http://www.bic.mni.mcgill.ca/ServicesSoftware/CIVET-2-1-0-Introduction}}.  CIVET corrects intensity non-uniformities using N3 \cite{sled1998nonparametric}\,; aligns the input volumes to the Talairach-like ICBM-152-nl template \cite{collins1994automatic}\,; classifies the image into white matter, gray matter, cerebrospinal fluid, and background \cite{zijdenbos2002automatic,tohka2004fast}\,; extracts the white-matter and pial surfaces \cite{kim2005automated}\,; and maps these to a common surface template \cite{lyttelton2007unbiased}. 

Cortical thickness was measured in native space at 81924 vertices using the Laplacian distance between the two surfaces. The Laplacian distance is the length of the path between the gray and white surfaces following the tangent vectors of the cortex represented as a Laplacian field \cite{jones2000three}. The CT measures were averaged into 78 regional measures relying on the AAL atlas. 

To extract the white/gray contrast measures, similarly to \cite{lewis2018t1}, the intensity on the T1-weighted MRI was sampled 1 mm inside and 1 mm outside of the white surface, and the ratio of the two measures was formed. Here we used a highly simplified version of the algorithm of \cite{lewis2018t1} and generated supra-white and sub-white surfaces relying on the surface normals provided by the CIVET pipeline. 

The intensity values on the T1-weighted image (without non-uniformity correction or normalization) were sampled at each vertex of both the supra-white surface and the sub-white surface, and

the ratio was formed by dividing the value at each vertex of the sub-white surface by the value at the corresponding vertex of the supra-white surface.

Similarly to CT measures, the contrast measures were averaged into 78 regional measures relying on the AAL atlas. 

The white/gray contrast measures are sensitive to scanner-specific differences in tissue contrast \cite{lewis2018t1}, so to correct for this, we normalized the contrast values per scanner manufacturer by  z-scoring the contrast values scanner manufacturer-wise as explained in detail in  \cite{lewis2018t1}. 

The CIVET pipeline failed to process $168$ subjects from the training set, $19$ subjects from the validation set and $19$ subjects from the test set most likely due to motion artifacts and/or excessive noise interfering with registration and segmentation. Consequently, a different model trained exclusively on the provided volumetric and sociodemographic data was used to infer the fluid intelligence score of the validation subjects whose derived data could not be produced.

\section{Machine learning approach}

The developed regression model based on artificial neural networks was trained with feature vectors that incorporated 122 volumetric, 78 contrast and 78 CT measures along with gender, age and the scanner manufacturer one-hot encoded. Age and image-derived attributes were normalized feature-wise by subtracting their mean and dividing them by their standard deviation. The network trained was a four-layer FNN. The model was trained using mini-batches of size 24 and Adam \cite{adam}, a stochastic gradient descent method with an adaptive learning rate starting from $\eta = 0.00001$. The cost function to minimize was the mean squared error (MSE). After every epoch the model was validated and the training stopped when the MSE was greater than the minimum MSE obtained in the previous iterations plus $0.7$. This stopping criteria was empirically found to increase the correlation coefficients of the predictions.

A four-layer FNN was trained by adjusting its parameters using the back-propagation algorithm to minimize the MSE produced between the desired output and the network prediction. The input layer consisted of 283 nodes corresponding to each of the features of the input vector. The two hidden layers have 20 and 15 nodes, respectively. Since predicting fluid intelligence scores is a single-output regression problem, the output layer consisted of a single node. As the network was fully connected, all nodes between successive layers were connected. This configuration provided the best results from the variations that were tried in the limited time available.

The weights of the FNN were randomly initialized as proposed in \cite{heinit} whereas the bias terms were initialized to zero. An exponential linear unit (ELU) nonlinear activation function \cite{ELU} was used in all intermediate layers with $\alpha=1$ such that if $x > 0$ then $f(x) = x$, otherwise $f(x) = exp(x) - 1$. All layers except for the input layer had a dropout of rate $0.5$ \cite{dropout}. The training of the FNN was implemented with TensorFlow \cite{tensorflow2015-whitepaper}.

\begin{figure}[h]
\centering
\includegraphics[width=0.75\textwidth]{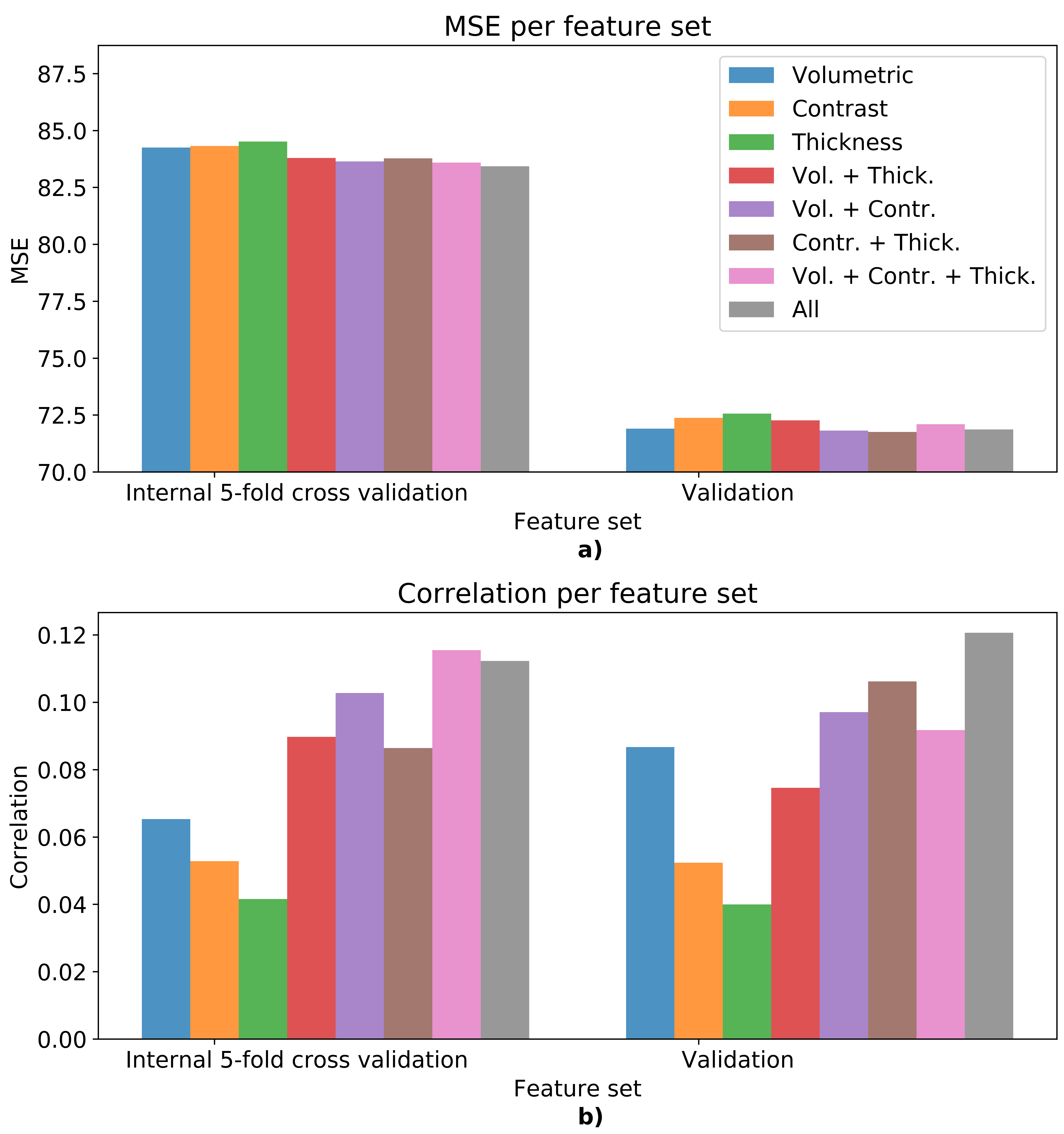}
\caption{Comparison of the MSE and correlation among FNNs trained on different feature sets in internal 5-fold cross-validation and validation set experiments. ``All" combines volumetric, contrast and thickness measures with age, gender and the scanner manufacturer.}
\label{feature_comparison}
\end{figure}

\section{Results}
Our predictions were generated on the regression model trained on the 3568 subjects for which all image-derived features were obtained. The final submission consisted of 4383 predictions achieving a MSE of 94.0270.

The performance of the model on the validation set and in an internal 5-fold cross-validation test was used to study how different combinations of features contribute to the MSE and correlation (Fig. \ref{feature_comparison}). The correlation here refers to Pearson correlation coefficient between the predicted and actual fluid intelligence scores. Both experiments showed that the combination of all features provided with larger correlation coefficients (Fig. \ref{feature_comparison} b). But, the variability of the MSE produced among different models trained on different set of features was small (Fig. \ref{feature_comparison} a).

\begin{figure}
\includegraphics[width=\textwidth]{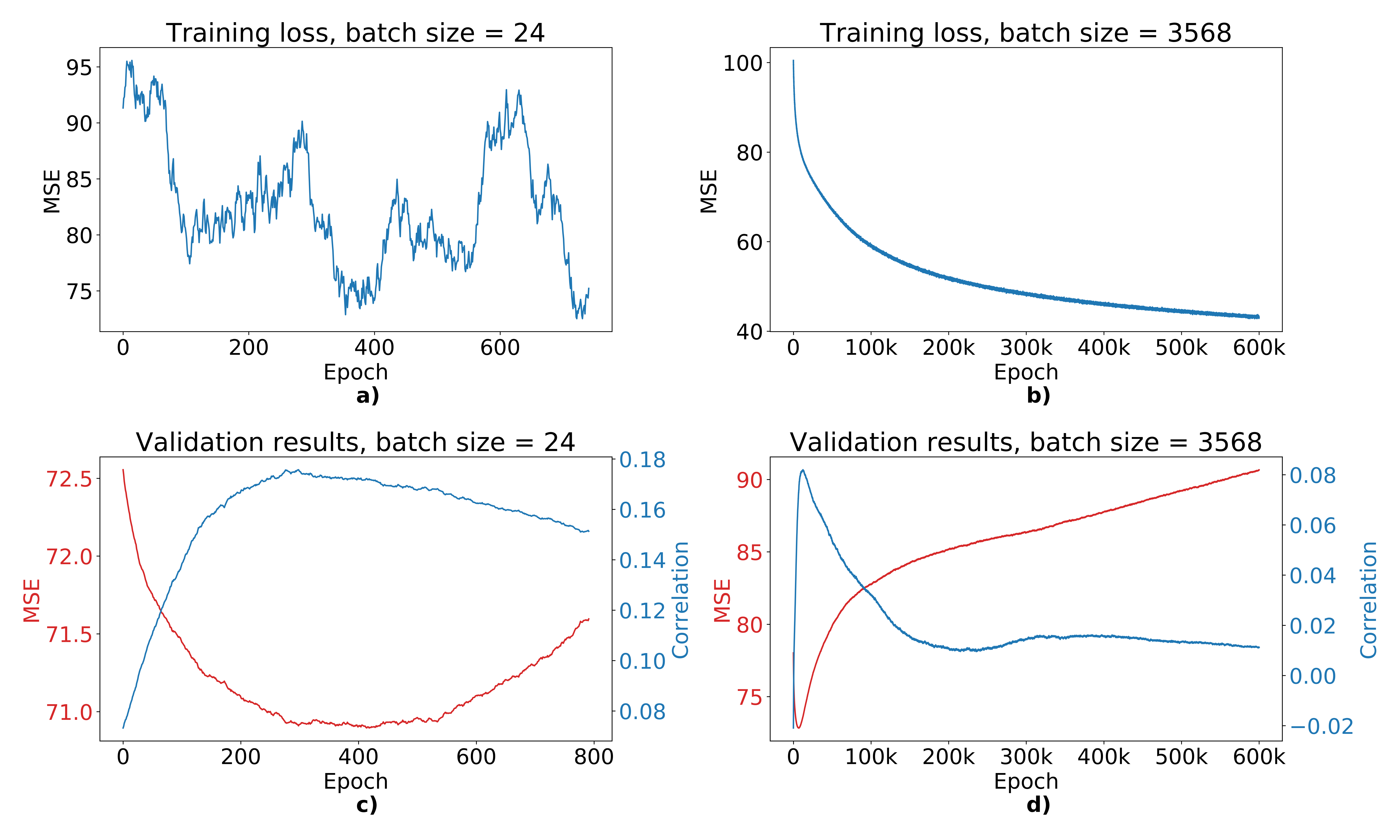}
\caption{Training loss and validation results of models trained with batch sizes of 24 and 3568 assessed on the validation set. Training loss is presented as moving average using a window size of 51 epochs.}
\label{validation_corr_mse_trainingloss}
\end{figure}

The proposed method trained with all features combined achieved a MSE of $81.89$ and a correlation of $0.13$ in an internal 5-fold cross-validation experiment. The results obtained on the validation set were a MSE of $71.596$ and a correlation of $0.151$, and its corresponding training loss was $84.28$ (Fig. \ref{validation_corr_mse_trainingloss} a)).

As depicted in Fig. \ref{validation_corr_mse_trainingloss} a) and b), choosing a small batch size to train the FNN causes large oscillations in the training loss whereas choosing the largest possible batch size leads to a steady decrease of the MSE. Providing with the entire training data set to train a neural network makes more accurate estimates of the gradient directions to minimize the loss over the training data. On the other hand, small batch sizes may need more iterations to converge, but the consequent fluctuations that occur during the training can lead to reaching other local minima with potentially better generalization capabilities. Fig. \ref{validation_corr_mse_trainingloss} c) and d) show that when the batch size is $3568$ not only the number of required epochs was significantly larger but also the proposed FNN provided with worse MSE and correlation than when the batch size is 24.

\vspace{-0.5in}

\subsection{Computation time}
The four-layer FNN was implemented in Tensorflow (Python). The total running time for training the model using all features was approximately 3 and a half minutes. Generating predictions of the testing set took half a second. The regression model was run on Ubuntu 16.04 with an Intel Xeon W-2125 CPU @ 4.00GHz processor and 64GB of memory.

Processing a subject through the CIVET pipeline entailed a computational time of approximately 10 hours on a cluster with Intel Xeon E5-2683 V4 @ 2.1 GHz processors using a single core with 4 GB of memory.

\section{Discussion}

We presented a method based on artificial neural networks to predict fluid intelligence scores from T1-weighted MR images, age and gender. Contrast and CT measures were additionally derived from the MR images to complement the provided volumetric and sociodemographic data. Training the proposed model with the combination of all image-derived features provided larger correlation coefficients than when it was trained solely on volumetric features. Nonetheless, the overall MSE of the predicted scores did not improve. The selected batch size to train the FNN caused the training loss to oscillate. However, as shown in Section 4, it increased the capability of the model to generalize. With this setting, the regression model achieved a MSE of $71.596$ and a correlation of $0.151$ in the validation set of 415 subjects. Due to the inherent complexity of the regression problem and the incorporation of additional image-derived features, future work is required to explore different architectures and deeper neural network models.

\section{Acknowledgements}

This research has been supported by grant 316258 from the Academy of Finland (to JT) and it is co-funded by Horizon 2020 Framework Programme of the European Union (Marie Skłodowska Curie grant agreement No 740264). The research also benefited from computational resources provided by Compute Canada (www.computecanada.ca) and Calcul Quebec (www.calculquebec.ca).

Data used in the preparation of this article were obtained from the Adolescent Brain Cognitive Development (ABCD) Study (\url{https://abcdstudy.org}), held in the NIMH Data Archive (NDA).  This is a multisite, longitudinal study designed to recruit more than 10,000 children age 9-10 and follow them over 10 years into early adulthood. The ABCD Study is supported by the National Institutes of Health and additional federal partners under award numbers U01DA041022, U01DA041025, U01DA041028, U01DA041048, U01DA041089, U01DA041093, U01DA041106, U01DA041117, U01DA041120, U01DA041134, U01DA041148, U01DA041156, U01DA041174, U24DA041123, and U24DA041147.  A full list of supporters is available at \url{https://abcdstudy.org/nih-collaborators}. A listing of participating sites and a complete listing of the study investigators can be found at \url{https://abcdstudy.org/principal-investigators.html}. ABCD consortium investigators designed and implemented the study and/or provided data but did not necessarily participate in analysis or writing of this report. This manuscript reflects the views of the authors and may not reflect the opinions or views of the NIH or ABCD consortium investigators.
 
 The ABCD data repository grows and changes over time. The ABCD data used in this report came from doi:10.15154/1503213 (Train set); doi:10.15154/1503306 (Validation set); doi:10.15154/1503307 (Test set).

 \bibliographystyle{splncs04}
 \bibliography{abcdbib}

\begin{thebibliography}{10}
\providecommand{\url}[1]{\texttt{#1}}
\providecommand{\urlprefix}{URL }
\providecommand{\doi}[1]{https://doi.org/#1}

\bibitem{tensorflow2015-whitepaper}
Abadi, M., et~al.: {TensorFlow}: Large-scale machine learning on heterogeneous
  systems (2015), \url{http://tensorflow.org/}, software available from
  tensorflow.org

\bibitem{NIH}
Akshoomoff, N., Beaumont, J., Bauer, P., et~al.: Nih toolbox cognition battery
  (cb): Composite scores of crystallized, fluid, and overall cognition.
  Monographs of the Society for Research in Child Development  \textbf{78}(4),
  119--132 (2013). \doi{10.1111/mono.12038},
  \url{https://onlinelibrary.wiley.com/doi/abs/10.1111/mono.12038}

\bibitem{ants}
Avants, B.B., Epstein, C.L., Grossman, M., Gee, J.C.: Symmetric diffeomorphic
  image registration with cross-correlation: Evaluating automated labeling of
  elderly and neurodegenerative brain. Medical image analysis  \textbf{12 1},
  26--41 (2008)

\bibitem{atropos}
Avants, B.B., Tustison, N.J., Wu, J., Cook, P.A., Gee, J.C.: An open source
  multivariate framework for n-tissue segmentation with evaluation on public
  data. Neuroinformatics  \textbf{9},  381--400 (2011)

\bibitem{carroll}
Carroll, J.: Human Cognitive Abilities: A Survey of Factor-Analytic Studies.
  Cambridge University Press., 1st edn. (1993). \doi{10.1017/CBO9780511571312}

\bibitem{ELU}
Clevert, D.A., Unterthiner, T., Hochreiter, S.: Fast and accurate deep network
  learning by exponential linear units (elus). arXiv  \textbf{abs/1511.07289}
  (2015)

\bibitem{collins1994automatic}
Collins, D.L., Neelin, P., Peters, T.M., Evans, A.C.: Automatic {3D}
  intersubject registration of {MR} volumetric data in standardized {T}alairach
  space. Journal of computer assisted tomography  \textbf{18}(2),  192--205
  (1994)

\bibitem{minimalprocessingpipeline}
Hagler, D.J., et~al.: Image processing and analysis methods for the adolescent
  brain cognitive development study. bioRxiv  (2018). \doi{10.1101/457739},
  \url{https://www.biorxiv.org/content/early/2018/11/04/457739}

\bibitem{heinit}
He, K., Zhang, X., Ren, S., Sun, J.: Delving deep into rectifiers: Surpassing
  human-level performance on imagenet classification. 2015 IEEE International
  Conference on Computer Vision (ICCV) pp. 1026--1034 (2015)

\bibitem{jones2000three}
Jones, S.E., Buchbinder, B.R., Aharon, I.: Three-dimensional mapping of
  cortical thickness using laplace's equation. Human brain mapping
  \textbf{11}(1),  12--32 (2000)

\bibitem{kim2005automated}
Kim, J.S., Singh, V., Lee, J.K., Lerch, J., Ad-Dab'bagh, Y., MacDonald, D.,
  Lee, J.M., Kim, S.I., Evans, A.C.: Automated {3-D} extraction and evaluation
  of the inner and outer cortical surfaces using a laplacian map and partial
  volume effect classification. Neuroimage  \textbf{27}(1),  210--221 (2005)

\bibitem{adam}
Kingma, D.P., Ba, J.: Adam: A method for stochastic optimization. arXiv
  \textbf{abs/1412.6980} (2015)

\bibitem{lewis2018t1}
Lewis, J.D., Evans, A.C., Tohka, J., Group, B.D.C., et~al.: T1 white/gray
  contrast as a predictor of chronological age, and an index of cognitive
  performance. NeuroImage  \textbf{173},  341--350 (2018)

\bibitem{lyttelton2007unbiased}
Lyttelton, O., Boucher, M., Robbins, S., Evans, A.: An unbiased iterative group
  registration template for cortical surface analysis. Neuroimage
  \textbf{34}(4),  1535--1544 (2007)

\bibitem{manjon2010adaptive}
Manj{\'o}n, J.V., Coup{\'e}, P., Mart{\'\i}-Bonmat{\'\i}, L., Collins, D.L.,
  Robles, M.: Adaptive non-local means denoising of mr images with spatially
  varying noise levels. Journal of Magnetic Resonance Imaging  \textbf{31}(1),
  192--203 (2010)

\bibitem{ncandapipeline}
Pfefferbaum, A., Kwon, D., Brumback, T., Thompson, W.K., Cummins, K., Tapert,
  S.F., Brown, S.A., Colrain, I.M., Baker, F.C., Prouty, D., Bellis, M.D.D.,
  Clark, D.B., Nagel, B.J., Chu, W., Park, S.H., Pohl, K.M., Sullivan, E.V.:
  Altered brain developmental trajectories in adolescents after initiating
  drinking. The American journal of psychiatry  \textbf{175 4},  370--380
  (2018)

\bibitem{SRI24}
Rohlfing, T., Zahr, N.M., Sullivan, E.V., Pfefferbaum, A.: The sri24
  multichannel atlas of normal adult human brain structure. Human brain mapping
   \textbf{31 5},  798--819 (2010)

\bibitem{sled1998nonparametric}
Sled, J.G., Zijdenbos, A.P., Evans, A.C.: A nonparametric method for automatic
  correction of intensity nonuniformity in {MRI} data. Medical Imaging, IEEE
  Transactions on  \textbf{17}(1),  87--97 (1998)

\bibitem{dropout}
Srivastava, N., Hinton, G., Krizhevsky, A., Sutskever, I., Salakhutdinov, R.:
  Dropout: A simple way to prevent neural networks from overfitting. Journal of
  Machine Learning Research  \textbf{15},  1929--1958 (2014),
  \url{http://jmlr.org/papers/v15/srivastava14a.html}

\bibitem{tohka2004fast}
Tohka, J., Zijdenbos, A., Evans, A.: Fast and robust parameter estimation for
  statistical partial volume models in brain {MRI}. Neuroimage  \textbf{23}(1),
   84--97 (2004)

\bibitem{tzourio2002automated}
Tzourio-Mazoyer, N., Landeau, B., Papathanassiou, D., Crivello, F., Etard, O.,
  Delcroix, N., Mazoyer, B., Joliot, M.: Automated anatomical labeling of
  activations in spm using a macroscopic anatomical parcellation of the {MNI}
  {MRI} single-subject brain. Neuroimage  \textbf{15}(1),  273--289 (2002)

\bibitem{zijdenbos2002automatic}
Zijdenbos, A.P., Forghani, R., Evans, A.C.: Automatic``pipeline" analysis of
  {3-D} {MRI} data for clinical trials: application to multiple sclerosis.
  Medical Imaging, IEEE Transactions on  \textbf{21}(10),  1280--1291 (2002)

\end{thebibliography}

\end{document}